\documentclass[twocolumn,showpacs,preprintnumbers,amsmath,amssymb,prb]{revtex4}
\usepackage{graphicx}
\usepackage{longtable}

\countdef\tube=8
\ifnum\tube<10 
\def\HeadDir{./}
\else
\def\HeadDir{./eps}
\fi
\begin{document}

\bibliographystyle{aip}

\setlength{\mathindent}{0.2cm}

\title{Dependence of exciton transition energy of single-walled carbon nanotubes on surrounding dielectric materials
}

%\author[TK]{Y.~Miyauchi},
%\author[TH]{R.~Saito},
%\author[TH]{K.~Sato},
%\author[NY]{Y.~Ohno},
%\author[NY]{S.~Iwasaki},
%\author[NY]{T.~Mizutani},
%\author[NCS]{J.~Jiang},
%\author[TK]{S.~Maruyama\corauthref{maru}}
%\corauth[maru]{Corresponding author. FAX: +81-3-5841-6421.}
%\ead{maruyama@photon.t.u-tokyo.ac.jp}
%\ead[url]{http://www.photon.t.u-tokyo.ac.jp/}
%
%\address[TK]{Department of Mechanical Engineering, The University of Tokyo, Tokyo 113-8656, Japan}
%\address[TH]{Department of Physics, Tohoku University and CREST, Sendai 980-8578, Japan}
%\address[NY]{Department of Quantum Engineering, Nagoya University, Nagoya 464-8603, Japan}
%\address[NCS]{Center for High Performance Simulation and Department of Physics, North Carolina State University, Raleigh, North Carolina 27695-7518, USA}

\author{Y. Miyauchi$^{a}$, R. Saito$^{b}$, K. Sato$^{b}$,
Y. Ohno$^{c}$, S. Iwasaki$^{c}$, T. Mizutani$^{c}$, J. Jiang$^{d}$,
S. Maruyama$^{a}$\footnote{Corresoponding author. FAX: +81-3-5841-6421. \\ \ E-mail: maruyama@photon.t.u-tokyo.ac.jp (S. Maruyama)}}

\affiliation{$^{a}$Department of Mechanical Engineering, The University of Tokyo, Tokyo 113-8656, Japan}
\affiliation{$^{b}$Department of Physics, Tohoku University and CREST, Sendai 980-8578, Japan}
\affiliation{$^{c}$Department of Quantum Engineering, Nagoya University, Nagoya 464-8603, Japan}
\affiliation{$^{d}$Center for High Performance Simulation and Department of Physics, North Carolina State University, Raleigh, North Carolina 27695-7518, USA}

\date{\today}
\begin{abstract}
We theoretically investigate the dependence of exciton transition
energies on dielectric constant of surrounding materials. We make a
simple model for the relation between dielectric constant of environment and a
static dielectric constant describing the effects of electrons in core
states, $\sigma$ bonds and surrounding materials. Although the
model is very simple, calculated results well reproduce experimental
transition energy dependence on dielectric constant of various
surrounding materials.
\end{abstract}
%\maketitle

%\begin{keyword}
%PL \sep nanotube \sep environment effect
%\PACS 78.30.-j \sep 78.30.Na \sep 78.30.Ly
%\end{keyword}
%\end{frontmatter}

\maketitle

% Ms number PACS numbers
% 81.05.Tp Fullerenes and related materials; diamonds, graphite

\section{Introduction}

Photoluminescence (PL) of single-walled carbon nanotubes (SWNTs) has
been intensively studied for elucidating their unusual optical and
electronic properties due to one
dimensionality \cite{oconnell02,weisman02sm,lebedkin03,resasco03,hartschuh03,MaruyamaCPL,lefebvre04b,fwang05,maultzsch05j,dukovic05,c1095,htoon05,plentz05,Ohno06-PRB-s,miyauchi06prb1,miyauchi06prb2}. Since
both of electron-electron repulsion and electron-hole binding energies
for SWNTs are considerably large compared with those for conventional
three-dimensional materials, the Coulomb interactions between
electron-electron and electron-hole play an important role in optical
transition of SWNTs
\cite{ando97a,zhao04prl,perebeinos04prl,spataru04,jiang07-ex,jiang07-opphmx}. Optical
transition energies of SWNTs are strongly affected by the change of
environment around SWNTs such as bundling \cite{a1119j}, surfactant
suspension \cite{lefebvre04b,Ohno06-PRB-s,moore03-surf} and DNA
wrapping \cite{strano06-DNA}. Lefebvre {\it et al.} \cite{lefebvre04b}
reported that the transition energies for suspended SWNTs between two
pillars fabricated by the MEMS technique are blue-shifted relative to
the transition energies for micelle-suspended SWNTs. Ohno {\it et
al.} \cite{Ohno06-PRB-s} have compared the PL of suspended SWNTs
directly grown on a grated quartz substrate using alcohol CVD technique
\cite{MaruyamaCPL} with SDS-wrapped SWNTs \cite{weisman02sm}. The
energy differences between air-suspended and SDS-wrapped SWNTs depend
on $(n, m)$ and type of SWNTs [type I ($(2n+m)$ mod 3 = 1) or type II
($(2n+m)$ mod 3 = 2)\cite{saitobook,a1093,w1089}].

Recently, Ohno {\it et al.} studied $E_{11}$ transition energies of SWNTs in
various surrounding materials with different dielectric constant,
$\kappa_{\rm env}$ \cite{Ohno07unp}. Observed dependence of $E_{11}$
on $\kappa_{\rm env}$ for a $(n,m)$ nanotube showed a tendency that
can be roughly expressed as
\begin{equation}
E_{11}=E_{11}^{\infty}+A_{nm}^{\rm exp}\kappa_{\rm env}^{-\alpha}
\label{eq:E11alpha}
\end{equation}
where $E_{11}^{\infty}$ denotes a transition energy when $\kappa_{\rm
env}$ is infinity, $A_{nm}^{\rm exp}$ is the maximum value of an
energy change of $E_{11}$ by $\kappa_{\rm env}$, and $\alpha$ is a
fitting coefficient in the order of 1, respectively. At this stage,
the reason why the experimental curve follows Eq.(\ref{eq:E11alpha})
is not clear.

In the previous theoretical studies of excitonic transition energies for
SWNTs \cite{ando97a,perebeinos04prl,jiang07-ex,jiang07-opphmx},
a screening effect of a surrounding material is mainly described using a
static dielectric constant $\kappa$. However, since $\kappa$ consists of both
$\kappa_{\rm env}$ and screening effect by nanotube itself,
$\kappa_{\rm tube}$, experimental dependence of transition energies on
dielectric constants of environment can not directly compared with
calculations \cite{ando97a,perebeinos04prl,jiang07-ex,jiang07-opphmx}
using the static dielectric constant $\kappa$. In this study, we make
a simple model for the relation between $\kappa_{\rm env}$ and
$\kappa$. The calculated results of excitons for different
$\kappa_{\rm env}$ reproduced well the experimental transition energy
dependence on dielectric constant of various surrounding materials.
 
\section{Theoretical method}
\subsection{Exciton transition energy}
 Within the extended tight-binding model
 \cite{jiang07-ex,jiang07-opphmx,w1089}, we calculated transition
 energies from the ground state to the first bright
 exciton state by solving the Bethe-Salpeter equation,
\begin{equation}
\begin{array}{llll}
& {\displaystyle \left\{[E({\bf k}_{\rm c})-E({\bf k}_{\rm
v})]\delta({\bf k}^{'}_{\rm c}, {\bf k}_{\rm c}) \delta({\bf
k}^{'}_{\rm v},{\bf k}_{\rm v})\right.  }\\ &{\displaystyle \left.
+K({\bf k}^{'}_{\rm c} {\bf k}^{'}_{\rm v},{\bf k}_{\rm c} {\bf
k}_{\rm v}) \right\} \Psi^{n}({\bf k}_{\rm c} {\bf k}_{\rm v})
=\Omega_{n}\Psi^{n}({\bf k}^{'}_{\rm c}{\bf k}^{'}_{\rm v}), }
\end{array}
\label{eq:BS}
\end{equation}
where ${\bf k}_{\rm c}$ and ${\bf k}_{\rm v}$ denote wave vectors of
the conduction and valence energy bands and $E({\bf k}_{\rm c})$ and
$E({\bf k}_{\rm v})$ are the quasi-electron and quasi-hole energies,
respectively. $\Omega_n$ is the energy of the $n$-th excitation of the
exciton $(n=0,1,2,\cdots)$, and $\Psi^{n}({\bf k}_{\rm c}{\bf k}_{\rm
v})$ are the excitonic wavefunctions. The kernel $K({\bf k}^{'}_{\rm
c} {\bf k}^{'}_{\rm v},{\bf k}_{\rm c} {\bf k}_{\rm v})$ describes the
Coulomb interaction between an electron and a hole. Details of the exciton
calculation procedure is the same as presented in Refs
\cite{jiang07-ex,jiang07-opphmx,w1089}.

The exciton wavefunction $|\Psi^{n}_{{\bf q}}>$ with a center-of-mass
momentum ${\bf q}(={\bf k}_c-{\bf k}_v)$ can be expressed as
\begin{equation}
|\Psi^{n}_{\bf q}>=\sum_{\bf k} Z^{n}_{{\bf k}c,({\bf k}-{\bf q})v}
c^{+}_{{\bf k}c}c_{({\bf k}-{\bf q})v}|0>
\label{exwf},
\end{equation}
where $Z^{n}_{{\bf k}c,({\bf k}-{\bf q})v}$ is the eigenvector of the
$n$-th $(n=0,1,2,\cdots)$ state of the Bethe-Salpeter equation, and
$|0\rangle$ is the ground state. Due to momentum conservation, the
photon-excited exciton is an exciton with ${\bf q}\approx0$ for parallel
excitations to the nanotube axis. In this Letter, we calculate the $n =
0$ state of ${\bf q} = 0$ exciton for each $(n,m)$ SWNT.

\subsection{Dielectric screening effect}
In our calculation, the unscreened Coulomb potential $V$ between
carbon $\pi$ orbitals is modeled by the Ohno potential
\cite{perebeinos04prl}. We consider the dielectric screening effect
within the random phase approximation (RPA). In the RPA, the static
screened Coulomb interaction $W$ is expressed as \cite{ando97a}
\begin{equation}
\begin{array}{llll}
W=V/\kappa\epsilon({\bf q}),
\end{array}
\label{cbi}
\end{equation} 
where $\epsilon({\bf q})$ is the dielectric function describing
effects of the polarization of the $\pi$ bands. $\kappa$ is a static
dielectric constant describing the effects of electrons in core
states, $\sigma$ bonds, and surrounding materials. In the calculation,
we directly calculate only the polarization for the $\pi$ band, and
the effects of electrons in core states, $\sigma$ bands, and
surrounding materials are represented by a single constant
$\kappa$. In the most accurate expression, the inhomogeneous and
nonlocal dielectric response of the nanotube itself and the
surrounding materials should be considered. However, it is not easy
within extended tight binding method. In this study, instead of
treating the complicated dielectric response including surrounding
materials, we make a simple model for a relation between the static
dielectric constant $\kappa$ and $\kappa_{\rm env}$ to obtain the
$E_{11}$ dependence on $\kappa_{\rm env}$.

\begin{figure}
\includegraphics[width=6cm,clip]{\HeadDir/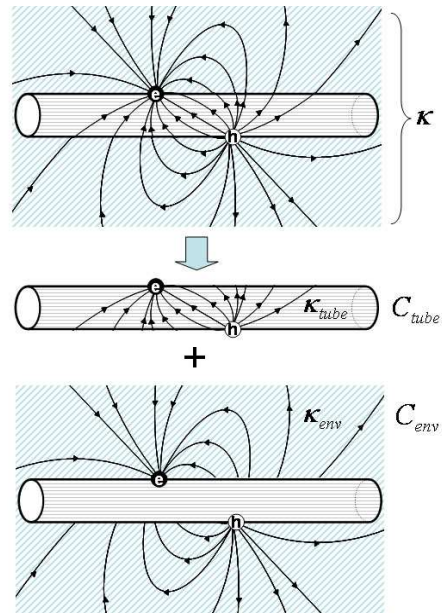}
\caption[]{Schematic of the connection of the net dielectric constant
$\kappa$ and the dielectric constant of the surrounding material
$\kappa_{\rm env}$ and the nanotube itself $\kappa_{\rm tube}$}.
\label{fig:1}
\end{figure}%

\subsection{Relationship between  $\kappa$ and $\kappa_{\rm env}$} 
Figure \ref{fig:1} shows a schematic view for the model
relationship between $\kappa$ and $\kappa_{\rm env}$. Here we consider
the screening effect related to $\kappa$ as a linear combination of
the screening of nanotube itself and the surrounding
material
\begin{equation}
\frac{1}{\kappa}=\frac{C_{\rm tube}}{\kappa_{\rm tube}}+\frac{C_{\rm env}}{\kappa_{\rm env}},
\label{eq:kapp-rel}
\end{equation}
where $\kappa_{\rm tube}$ is the dielectric constant within a nanotube
except for the $\pi$ bands, and $C_{\rm tube}$ and $C_{\rm env}$ are
coefficients for the inside and outside of a nanotube,
respectively. As shown in Eq.(\ref {eq:E11alpha}), the transition
energies observed in the experiment \cite{Ohno07unp} indicate that
there is a limit value \cite{perebeinos04prl} when $\kappa_{\rm env}$
$\rightarrow$ $\infty$. Hence, when $\kappa_{\rm env}$ $\rightarrow$
$\infty$, $C_{\rm env}/\kappa_{\rm env}$ can be removed from
Eq.(\ref{eq:kapp-rel}), and $1/\kappa$ is expressed by the limit value as
\begin{equation}
\frac{1}{\kappa} = \frac{C_{\rm tube}}{\kappa_{\rm tube}} \equiv \frac{1}{\kappa^{\infty}_{\rm tube}}, (\kappa_{\rm env} \rightarrow \infty )
\end{equation}
where $\kappa^{\infty}_{\rm tube}$ is the limit value of the net
dielectric constant $\kappa$ when $\kappa_{\rm env}$ is
infinity. Since electric flux lines through inside of the nanotube
remain even when $\kappa_{\rm env}$ $\rightarrow$ $\infty$, we assume
there is a certain value of $\kappa$ ($\kappa^{\infty}_{\rm tube}$)
that corresponds to the situation when dielectric screening by
surrounding material is perfect and only dielectric response of the
nanotube itself contributes to the net screening effect.

\begin{figure}
\includegraphics[width=6cm,clip]{\HeadDir/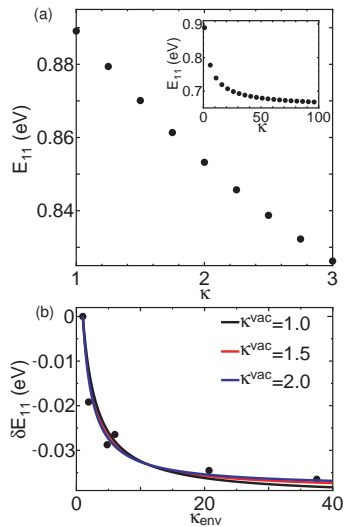}
\caption[]{(a) The $E_{11}$ energy for a $(9,8)$ SWNT as a function of
$\kappa$. (b) $\delta E_{11}$ dependence on $\kappa_{\rm env}$. Inset
in (a) shows the $E_{11}$ dependence up to $\kappa=100$. In (b),
circles denote the experimental data and solid curves denote the
calculated results of Eq.(\ref{eq:kenvdep}) for $\kappa^{\rm vac}=1.0$
(black), $1.5$ (red) and $2.0$ (blue).}
\label{fig:2}
\end{figure}

Replacing $C_{\rm tube}/\kappa_{\rm tube}$ by $\kappa^{\infty}_{\rm
tube}$, Eq.(\ref {eq:kapp-rel}) is modified as
\begin{equation}
\frac{1}{\kappa}=\frac{1}{\kappa^{\infty}_{\rm tube}}+\frac{C_{\rm env}}{\kappa_{\rm env}}.
\label{kapp-rel-mod}
\end{equation}
Next, we imagine that the SWNT is placed in the vacuum, which
corresponds to $\kappa=\kappa^{\rm vac}$ and $\kappa_{\rm env}=1$, and then
$C_{\rm env}$ can be expressed as
\begin{equation}
C_{\rm env}=\frac{1}{\kappa^{\rm vac}}-\frac{1}{\kappa^{\infty}_{\rm tube}},
\end{equation}
where $\kappa^{\rm vac}$ is the static dielectric constant {\it not}
for the vacuum, but for the situation that the nanotube is placed in
the vacuum. We now express $\kappa$ as a function of $\kappa_{\rm
env}$ through two parameters $\kappa^{\infty}_{\rm tube}$ and
$\kappa^{\rm vac}$, whose values can be estimated from the following
discussions. In the previous papers
\cite{ando97a,perebeinos04prl,jiang07-ex,jiang07-opphmx}, $\kappa$
value is put around 2 to obtain a good fit with experiments for
SWNTs with surrounding materials. Jiang {\it et al.}
\cite{jiang07-ex} have compared the calculated results with the
results for the two photon absorption experiments \cite{dukovic05}, and
obtained the best fit using $\kappa=2.22$ for SWNTs in a polymer
matrix. Here, since $\kappa^{\rm vac}$ is for nanotubes without
surrounding materials, $\kappa^{\rm vac}$ should be less than about 2
and close to 1 due to vacancy of inside of the tubes. With regard to
$\kappa^{\infty}_{\rm tube}$, according to the experimental results
\cite{lefebvre04b,Ohno06-PRB-s,Ohno07unp}, transition energy change
due to change of surrounding materials is at most 30-100meV.
Fig.\ref{fig:2}(a) shows the calculated $E_{11}$ energy dependence on
$\kappa$ for a (9,8) SWNT in a small $\kappa$ region, while the inset shows
the $E_{11}$ dependence up to $\kappa_{\rm env}=100$. As shown in
Fig.\ref{fig:2}(a), variation of $\kappa$ that yields the transition
energy change of 30 to 100 meV is about 1 to 3 when $\kappa$ is around
2. Therefore, the value of $\kappa^{\infty}_{\rm tube}$ should be
around 2 to 3 and that of $\kappa^{\rm vac}$ should be around 1 to 2.

\subsection{Dependence of excitation energy on $\kappa_{\rm env}$}
 
As shown in Fig.\ref{fig:2}(a), the calculated $E_{11}$ energies
decrease with increasing $\kappa$. This is mainly due to the fact that
the self energy (e-e repulsion) always exceeds to the e-h binding
energy and that the both interactions (e-e and e-h) decrease with
increasing $\kappa$. The $E_{11}$ almost linearly depends on $\kappa$
around the small $\kappa$ region. We checked that the linear
dependence is universal for all $(n,m)$'s for diameters more than 0.7
nm. Assuming the linear dependence, variation of the excitation energy
$\delta E_{11} \equiv E_{11}-E_{11}(\kappa_{\rm env}=1)$ for the small
$\kappa$ region is approximated by
\begin{equation}
\delta E_{\rm 11}=-A_{nm}(\kappa-\kappa^{\rm vac}),
\label{eq:keffdep}
\end{equation}
where $A_{\rm nm}$ is the gradient of $\delta E_{\rm 11}$ near the
small $\kappa$ region for each $(n,m)$ type. After we transform
$\kappa$ using the relationship of Eq.(\ref{eq:kapp-rel}), 
Eq.(\ref{eq:keffdep}) is modified as
\begin{equation}
\begin{array}{llll}
{\displaystyle
\delta E_{\rm 11}=-A_{nm}(\kappa^{\infty}_{\rm tube}-\kappa^{\rm
vac})(\frac{\kappa_{\rm env}-1}{\kappa_{\rm
env}+(\kappa^{\infty}_{\rm tube}-\kappa^{\rm vac})/\kappa^{\rm vac}}).
}
\label{eq:kenvdep}
\end{array}
\end{equation}

$A_{nm}(\kappa^{\infty}_{\rm tube}-\kappa^{\rm vac})$ corresponds to
the maximum value of $\delta E_{\rm 11}$ when $\kappa_{\rm env}$
$\rightarrow$ $\infty$, which corresponds to the value of coefficient
$A_{nm}^{\rm exp}$ in the fitting curve of Eq.(\ref{eq:E11alpha}).  For
$(9,8)$ SWNT, the fitted value to the calculated results for $A_{nm}$
is 33 meV and $A_{nm}^{\rm exp}$ obtained by the fit to the
experiment\cite{Ohno07unp} using Eq.(\ref{eq:E11alpha}) is 36 meV,
and $\kappa^{\infty}_{\rm tube}-\kappa^{\rm vac}$ should be around 1. 
The values for $\kappa^{\infty}_{\rm tube}$ and $\kappa^{\rm vac}$
are consistent with the values conventionally used for SWNTs in
dielectric materials
\cite{ando97a,perebeinos04prl,jiang07-ex,jiang07-opphmx}.

\section{Results and Discussion}
Figure \ref{fig:2}(b) compares $\delta E_{11}$ for a (9,8) SWNT
depending on $\kappa_{\rm env}$ by the experiment (solid circles) and
the calculated results (lines) for $\kappa^{\rm vac}=1,1.5,2.0$ using
Eq.(\ref{eq:kenvdep}). As shown in Fig.\ref{fig:2}(b), the qualitative
shape of theoretical curves are in good agreement with the experiment
and not affected so much by the change of $\kappa^{\rm vac}$. Since
the exact value of $\kappa_{\rm vac}$ is unknown, we hereafter set
$\kappa^{\rm vac}=1.5$ for each $(n,m)$ SWNT. For the $(9,8)$ SWNT in
Fig.\ref{fig:2}(b), $\kappa^{\infty}_{\rm tube}=2.7$ and $\kappa^{\rm
vac}=1.5$ are fitting values. These values are consistent with the
discussion in the previous section.  After setting $\kappa^{\rm
vac}=1.5$, Eq.(\ref{eq:kenvdep}) turns to be
\begin{equation}
\begin{array}{llll}
{\displaystyle
\delta E_{\rm 11}=\frac{-A_{nm}(\kappa_{\rm env}-1)}{\kappa_{\rm
env}/(\kappa^{\infty}_{\rm tube}-\kappa^{\rm vac})+1/1.5}.
\label{eq:kenvdep2}
}
\end{array}
\end{equation}

Thus, we express $\delta E_{\rm 11}$ as a function of $\kappa_{\rm
env}$ with one parameter $(\kappa^{\infty}_{\rm tube}-\kappa^{\rm
vac})$.

\begin{figure}
\includegraphics[width=5cm,clip]{\HeadDir/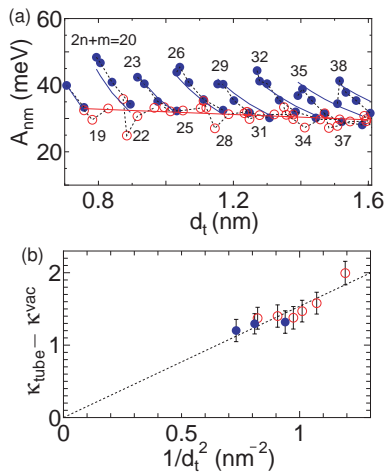}
\caption[]{(a) Calculated values of $A_{nm}$ for each $(n,m)$
SWNT. Open (red) and solid (blue) circles correspond to type I and
type II SWNTs, respectively. Solid lines denote the fit curve by
Eq.(\ref{eq:Anm}). (b) $\kappa^{\infty}_{\rm tube}-\kappa^{\rm vac}$
vs $1/d_t^2$. The values of $\kappa^{\infty}_{\rm tube}-\kappa^{\rm
vac}$ are obtained by the fit of Eq.(\ref{eq:kenvdep}) to the
experimental data for each $(n,m)$.}
\label{fig:3}
\end{figure}%

Figure \ref{fig:3}(a) shows the calculated values of $A_{nm}$ for each
$(n,m)$'s. Family pattern of $(2n+m=const.)$ family is drawn with the
$2n+m$ values by dotted lines. We found a slight diameter dependence
and relatively large chiral angle dependence of $A_{nm}$ for type II
SWNTs (blue) compared with type I SWNTs (red). The type II SWNTs with
larger chiral angles tend to have larger value of $A_{nm}$. For a
convenient use of Eq.(\ref{eq:kenvdep}), we give a fitting function of
$A_{nm}$ meV as
\begin{equation}
A_{nm}=A+Bd_t+(C+D/d_t)\cos{3\theta},
\label{eq:Anm}
\end{equation} 
which gives the average (maximum) error of $\pm 2\rm meV$$(8 \rm meV)$
for type I, and $\pm 2\rm meV$$(5 \rm meV)$ for type II SWNTs. The fit
curve is shown in Fig.\ref{fig:3}(a) by solid lines. Here $d_t$ (nm)
is the diameter of nanotube and $\theta$ is the chiral
angle \cite{saitobook}. The values of (A, B, C, D) are (36, -4, 0, 0)
and (33, -3, 6, 7) for type I and for type II SWNTs, respectively.

In order to expand our result to many $(n,m)$ SWNTs, we need a
function to describe $(\kappa^{\infty}_{\rm tube}-\kappa^{\rm
vac})$. It is important to note that $\kappa^{\infty}_{\rm tube}$
should depend on the diameter. An exact function should be calculated
by taking into account the Coulomb interaction considering induced
surface charge at the boundary of the nanotube and surrounding materal
for an e-e or e-h pair for each $(n,m)$ SWNT. Instead of calculating
this complicated function, here we roughly estimate the
$(\kappa^{\infty}_{\rm tube}-\kappa^{\rm vac})$ as a simple function
of diameter $d_t$, since $(\kappa^{\infty}_{\rm tube}-\kappa^{\rm
vac})$ should depend on the cross section of a SWNT. As shown in
Fig.\ref{fig:3}(b), $(\kappa^{\infty}_{\rm tube}-\kappa^{\rm vac})$ is
roughly proportional to $1/d_t^2$,
\begin{equation}
(\kappa^{\infty}_{\rm tube}-\kappa^{\rm vac})=\frac{E}{d_t^2},
\label{eq:Bdt2}
\end{equation}
with the coefficient $E=1.5\pm0.3$ $\rm nm^{2}$. Here $(\kappa^{\infty}_{\rm
tube}-\kappa^{\rm vac})$ is obtained by the fit using
Eq.(\ref{eq:kenvdep}) and $A_{nm}$ calculated for each
chirality. Fig.\ref{fig:3}(b) clearly shows that our calculated
$A_{nm}$ well describes the chiral angle dependence of $\delta E_{11}$
and that the remaining diameter dependence is understood by
$(\kappa^{\infty}_{\rm tube}-\kappa^{\rm vac})$ through $1/d_t^2$. This
$1/d_t^2$ dependence implies that $\kappa^{\infty}_{\rm tube}$ depends
on the volume of inner space of the nanotube. Although the number of
experimental data available for the fit is small and selection of this
function is arbitrary to some extent, it is reasonable that
$1/\kappa^{\infty}_{\rm tube}$ increase with the increase of the
diameter, since $1/\kappa^{\infty}_{\rm tube}$ corresponds to the
Coulomb interaction through the inner space of the nanotube. In order
to find an accurate form of the function, future experiments and
theoretical studies are definitely needed.

\begin{figure}
\includegraphics[width=9cm,clip]{\HeadDir/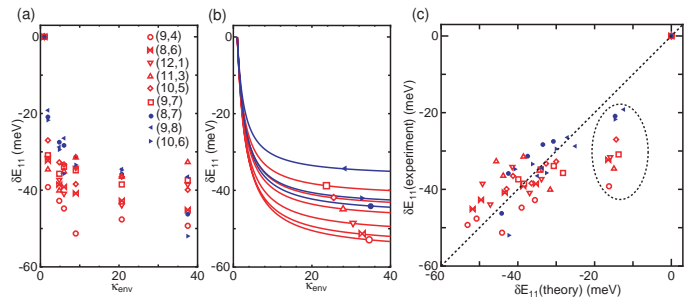}
\caption[]{The transition energy dependence plotted as a function of
$\kappa_{\rm env}$. (a) experiment and (b) calculated results are
indicated by (a) symbols and (b) solid curves. In (b), $(n,m)$ for
each curve is indicated by a symbol on the curve. (c) Comparison of
$\delta E_{11}$ for the experiment ($\delta E_{11}$(experiment)) and
calculation ($\delta E_{11}$(theory)). A dotted line indicates the line
of $\delta E_{11}\rm (experiment)=\delta E_{11}\rm (theory)$. Open
(red) and solid (blue) symbols correspond to type I and type II SWNTs,
respectively. The data in the dotted circle are the data for
$\kappa_{\rm env}=1.9$ \cite{Ohno07unp} (see text).}
\label{fig:4}
\end{figure}%

Figure \ref{fig:4} shows $\delta E_{11}$ as a function of $\kappa_{\rm
env}$ for (a) the experiment and (b) the calculation using
Eq.(\ref{eq:kenvdep2}) and (\ref{eq:Bdt2}). Fig.\ref{fig:4}(c)
compares $\delta E_{11}$ for the experiment and that for the
calculation with the same $\kappa_{\rm env}$ values. The same symbols
for an $(n,m)$ are used in three figures of Fig.\ref{fig:4}. Details
of experimental data will be published elsewhere
\cite{Ohno07unp}. Although our treatment is very simple, the
calculated curves for various $(n,m)$ SWNTs well reproduce the
experimentally observed tendency for each $(n,m)$ SWNT, and the degree
of difference between each $(n,m)$ type is also in good agreement with
the experiment. As shown in Fig.\ref{fig:4}(c), $\delta E_{11}\rm
(theory)$ is in a good agreement with $\delta E_{11}\rm (experiment)$
except for several points indicated by a dotted circle in the figure,
which correspond to a case for the smallest $\kappa_{\rm env}=1.9$
(hexane) except for $\kappa_{\rm env}=1$ (air) in the experimental data
\cite{Ohno07unp}. The value of $\kappa_{\rm env}=1.9$ for hexane is
adopted as the dielectric constant for the material, in which the
dipole moments of liquid hexane are not aligned perfectly even in the
presence of the electric field. Since $\kappa_{\rm env}=1.9$ is a
macroscopic value, a local dielectric response might be different from
the averaged macroscopic response. If the local dielectric constant
near SWNTs becomes large (for example, $\kappa_{\rm env}\approx 3$),
the fitting of Fig.\ref{fig:4}(c) becomes better. We expect that the
dipole moments of a dielectric material might be aligned locally for a
strong electric field near an exciton, which makes the local
dielectric constant relatively large. This will be an interesting
subject for exciton PL physics. Since the difference of $A_{nm}$
between each $(n,m)$ type decreases with increasing the diameter, it
is predicted that the amount of variation due to the change of
$\kappa_{\rm env}$ mostly depend on diameter in the larger diameter
range. Thus a PL experiment for nanotubes with large diameters would
be desirable for a further comparison.

\section{Summary}

 In summary, the dependence of exciton transition energies on
 dielectric constant of surrounding materials are investigated. We
 proposed a model for the relation between dielectric constant of the
 environment and a static dielectric constant $\kappa$ in the
 calculation. Although the model is quite simple, calculated results
 well reproduce the feature of experimentally observed transition
 energy dependence on dielectric constant of various surrounding
 materials, and various $d_t$ and $\theta$.

\section*{Acknowledgments}
Y.M. is supported by JSPS Research Fellowships for Young Scientists
(No. \ 16-11409). R.S. acknowledges a Grant-in-Aid (No.\ 16076201) from
the Ministry of Education, Japan.

%\begin{figure}[p] %with [p], you can put figures at the end of the manuscript
%\includegraphics[width=10cm,clip]{\HeadDir/fig1.eps}
%\caption[]{Schematic of the connection of the net dielectric constant
%$\kappa$ and the dielectric constant of the surrounding material
%$\kappa_{\rm env}$ and the nanotube itself $\kappa_{\rm tube}$}.
%\label{fig:1}
%\end{figure}%

\end{document}